\documentclass[conference, 10pt]{IEEEtran}
\IEEEoverridecommandlockouts
\usepackage{cite}
\usepackage{amsmath,amssymb,amsfonts}
\usepackage{algorithmic}
\usepackage{graphicx}
\usepackage{textcomp}
\usepackage[dvipsnames]{xcolor}
\usepackage{tikz}
\usepackage{pgfplots}
\usepackage{hyperref}
\usepackage{xfrac}
\usepackage{tablefootnote}

\pgfplotsset{compat=1.18} 

\def\BibTeX{{\rm B\kern-.05em{\sc i\kern-.025em b}\kern-.08em
    T\kern-.1667em\lower.7ex\hbox{E}\kern-.125emX}}
    
\begin{document}
\newcommand{\bytes}{\mbox{bytes}}
\newcommand{\byte}{\mbox{byte}}
\newcommand{\second}{\mbox{s}}
\newcommand{\seconds}{\mbox{s}}
\newcommand{\flop}{\mbox{flop}}
\newcommand{\flops}{\mbox{flops}}
\newcommand{\NJFLOP}{\mbox{nJ/\flop}}
\newcommand{\instr}{\mbox{instr}}
\newcommand{\cycle}{\mbox{cy}}
\newcommand{\iter}{\mbox{it}}
\newcommand{\cycles}{\mbox{cy}}
\newcommand{\FCY}{\mbox{\flop/\cycle}}
\newcommand{\FIT}{\mbox{\flop/\iter}}
\newcommand{\BIT}{\mbox{\byte/\iter}}
\newcommand{\FR}{\mbox{\flops/\mbox{row}}}
\newcommand{\BR}{\mbox{\byte/\mbox{row}}}
\newcommand{\CYF}{\mbox{\cycles/\flop}}
\newcommand{\CS}{\mbox{\cycle/\second}}
\newcommand{\GCS}{\mbox{G\cycle/\second}}
\newcommand{\word}{\mbox{Word}}
\newcommand{\words}{\mbox{Words}}
\newcommand{\order}[1]{\mbox{${\cal O}\left(\mbox{#1}\right)$}}
\newcommand{\bit}{\mbox{bit}}
\newcommand{\bits}{\mbox{bits}}
\newcommand{\GBPS}{\mbox{G\bit/\second}}
\newcommand{\MBPS}{\mbox{M\bit/\second}}
\newcommand{\FS}{\mbox{\flop/\second}}
\newcommand{\BS}{\mbox{\byte/\second}}
\newcommand{\BC}{\mbox{\byte/\cycle}}
\newcommand{\GBS}{\mbox{G\byte/\second}}
\newcommand{\MBS}{\mbox{M\byte/\second}}
\newcommand{\GWS}{\mbox{G\word/\second}}
\newcommand{\GFS}{\mbox{G\flop/\second}}
\newcommand{\MFS}{\mbox{M\flop/\second}}
\newcommand{\lup}{\mbox{LUP}}
\newcommand{\lups}{\mbox{LUPs}}
\newcommand{\LUPCY}{\mbox{\lup/\cycle}}
\newcommand{\LUPS}{\mbox{\lup/\second}}
\newcommand{\MLUPS}{\mbox{M\lup/\second}}
\newcommand{\GLUPS}{\mbox{G\lup/\second}}
\newcommand{\GHZ}{\mbox{GHz}}
\newcommand{\ns}{\mbox{ns}}
\newcommand{\WF}{\mbox{\word/\flop}}
\newcommand{\BF}{\mbox{\byte/\flop}}
\newcommand{\FB}{\mbox{\flop/\byte}}
\newcommand{\BL}{\mbox{\byte/\lup}}
\newcommand{\GB}{\mbox{GB}}
\newcommand{\KB}{\mbox{kB}}
\newcommand{\MB}{\mbox{MB}}
\newcommand{\GiB}{\mbox{GiB}}
\newcommand{\MiB}{\mbox{MiB}}
\newcommand{\KiB}{\mbox{KiB}}
\newcommand{\TB}{\mbox{TB}}
\newcommand{\TiB}{\mbox{TiB}}
\newcommand{\W}{\mbox{W}}
\newcommand{\muarch}{\mbox{$\mu$-arch}}
\newcommand{\muop}{\mbox{$\mu$-op}}
\newcommand{\muops}{\mbox{$\mu$-ops}}

\newcommand{\eos}{\;.}
\newcommand{\cma}{\;,}
\newcommand{\rlm}{roof{}line model}
\newcommand{\rl}{roof{}line}
\newcommand{\Rlm}{Roof{}line model}
\newcommand{\Rl}{Roof{}line}
\newcommand{\olsep}{\|}
\newcommand{\nolsep}{|}
\newcommand{\ecmspace}{\,}
\newcommand{\TOL}{$T_{\mathrm{c}\_\mathrm{OL}}$}
\newcommand{\NNZR}{$N_\mathrm{nzr}$}
\newcommand{\NR}{$N_\mathrm{r}$}
\newcommand{\NNZ}{$N_\mathrm{nz}$}
\newcommand{\ecm}[6]{\mbox{$\left\{{#1}\ecmspace\olsep\ecmspace {#2}\ecmspace\nolsep\ecmspace {#3}\ecmspace\nolsep\ecmspace {#4}\ecmspace\nolsep\ecmspace {#5}\right\}\ecmspace{#6}$}}
\newcommand{\epsep}{\rceil}
\newcommand{\ecmp}[4]{\mbox{$\left\{{#1}\ecmspace\epsep\ecmspace {#2}\ecmspace\epsep\ecmspace {#3}\right\}\ecmspace{#4}$}}
\newcommand{\ecme}[4]{\mbox{$\left({#1}\ecmspace\epsep\ecmspace {#2}\ecmspace\epsep\ecmspace {#3}\right)\ecmspace{#4}$}}
\newcommand{\sellcs}{SELL-\texorpdfstring{$C$-$\sigma$}{C-sigma}}
\newcommand{\likwid}{\texttt{LIKWID}}
\newcommand{\likwidperfctr}{\texttt{likwid-perfctr}}
\newcommand{\likwidpin}{\texttt{likwid-pin}}
\newcommand{\likwidbench}{\texttt{likwid-bench}}
\newcommand{\lmbench}{\texttt{lmbench}}
\newcommand{\afx}{A64FX}
\newcommand{\spmv}{SpMV}
\newcommand{\cmg}{CMG}
\newcommand{\mve}{MVE}
\newcommand{\crs}{CRS}
\newcommand{\ellpack}{ELLPACK}
\newcommand{\tands}{dRECT}
\newcommand{\bq}{\begin{equation}}
\newcommand{\eq}{\end{equation}}

\newcommand{\inputtikz}[2]{%
	\input{#1/#2.tex}%
}

\newenvironment{customlegend}[1][]{%
	\begingroup
	\csname pgfplots@init@cleared@structures\endcsname
	\pgfplotsset{#1}%
}{%
	\csname pgfplots@createlegend\endcsname
	\endgroup
}%
\def\addlegendimage{\csname pgfplots@addlegendimage\endcsname}

\makeatletter
\newcommand{\linebreakand}{%
  \end{@IEEEauthorhalign}
  \hfill\mbox{}\par
  \mbox{}\hfill\begin{@IEEEauthorhalign}
}
\makeatother

\newcommand{\GHacomm}[1]{{\color{red}-- {#1} --\color{black}} }
\newcommand{\JLcomm}[1]{{\color{orange}-- {#1} --\color{black}} }
\newcommand{\DOcomm}[1]{{\color{magenta}-- {#1} --\color{black}} }
\newcommand{\TGcomm}[1]{{\color{tumbleweed}-- {#1} --\color{black}} }
\newcommand{\revon}{\color{black}}
\newcommand{\revoff}{\color{black}}


\definecolor{amber}{rgb}{1.0, 0.75, 0.0}
\definecolor{amethyst}{rgb}{0.6, 0.4, 0.8}
\definecolor{applegreen}{rgb}{0.55, 0.71, 0.0}
\definecolor{tumbleweed}{rgb}{0.87, 0.67, 0.53}
\definecolor{greyback}{RGB}{248,248,248}
\definecolor{deepblue}{RGB}{43,131,186}
\definecolor{greenalt}{RGB}{35,139,69}
\definecolor{darkred}{RGB}{215,25,28}
\definecolor{darkorange}{RGB}{253,174,97}
\definecolor{light-gray}{gray}{0.85}
\definecolor{lighter-gray}{gray}{0.95}
\definecolor{applegreen}{rgb}{0.55, 0.71, 0.0}
\definecolor{asparagus}{rgb}{0.53, 0.66, 0.42}
\definecolor{babyblueeyes}{rgb}{0.63, 0.79, 0.95}
\definecolor{burntsienna}{rgb}{0.91, 0.45, 0.32}
\definecolor{deepcarmine}{rgb}{0.66, 0.13, 0.24}
\definecolor{lightcornflowerblue}{rgb}{0.6, 0.81, 0.93}
\definecolor{steelblue}{rgb}{0.27, 0.51, 0.71}
\definecolor{pastelblue}{rgb}{0.68, 0.78, 0.81}
\definecolor{pastelbrown}{rgb}{0.51, 0.41, 0.33}
\definecolor{pastelgray}{rgb}{0.81, 0.81, 0.77}
\definecolor{pastelgreen}{rgb}{0.47, 0.87, 0.47}
\definecolor{pastelmagenta}{rgb}{0.96, 0.6, 0.76}
\definecolor{pastelorange}{rgb}{1.0, 0.7, 0.28}
\definecolor{pastelpink}{rgb}{1.0, 0.82, 0.86}
\definecolor{pastelpurple}{rgb}{0.7, 0.62, 0.71}
\definecolor{pastelred}{rgb}{1.0, 0.41, 0.38}
\definecolor{pastelviolet}{rgb}{0.8, 0.6, 0.79}
\definecolor{pastelyellow}{rgb}{0.99, 0.99, 0.59}
\definecolor{darkblue}{rgb}{0.0, 0.0, 0.55}

\definecolor{blue-seq-0}{RGB}{255, 255, 204}
\definecolor{blue-seq-1}{RGB}{119, 233, 180}
\definecolor{blue-seq-2}{RGB}{127, 205, 187}
\definecolor{blue-seq-3}{RGB}{ 65, 182, 196}
\definecolor{blue-seq-4}{RGB}{ 44, 127, 184}
\definecolor{blue-seq-5}{RGB}{ 37,  52, 148}
\definecolor{red-seq-0}{RGB}{254, 229, 217}
\definecolor{red-seq-1}{RGB}{252, 187, 161}
\definecolor{red-seq-2}{RGB}{252, 146, 114}
\definecolor{red-seq-3}{RGB}{251, 106, 74}
\definecolor{red-seq-4}{RGB}{222,  45, 38}
\definecolor{red-seq-5}{RGB}{165,  15, 21}
\definecolor{green-seq-0}{RGB}{237,248,233}
\definecolor{green-seq-1}{RGB}{199,233,192}
\definecolor{green-seq-2}{RGB}{161,217,155}
\definecolor{green-seq-3}{RGB}{116,196,118}
\definecolor{green-seq-4}{RGB}{49,163,84}
\definecolor{green-seq-5}{RGB}{0,109,44}
\definecolor{qual-0}{RGB}{102,194,165}
\definecolor{qual-1}{RGB}{252,141,98}
\definecolor{qual-2}{RGB}{141,160,203}
\definecolor{qual-3}{RGB}{231,138,195}
\definecolor{qual-4}{RGB}{166,216,84}
\definecolor{qual-5}{RGB}{255,217,47}
\definecolor{qual-6}{RGB}{229,196,148}
\definecolor{cbf-0}{RGB}{215,48,39}
\definecolor{cbf-1}{RGB}{244,109,67}
\definecolor{cbf-2}{RGB}{253,174,97}
\definecolor{cbf-3}{RGB}{254,224,144}
\definecolor{cbf-4}{RGB}{224,243,248}
\definecolor{cbf-5}{RGB}{171,217,233}
\definecolor{cbf-6}{RGB}{116,173,209}
\definecolor{cbf-7}{RGB}{69,117,180}

\definecolor{longqual-0}{RGB}{166,206,227}
\definecolor{longqual-1}{RGB}{31,120,180}
\definecolor{longqual-2}{RGB}{178,223,138}
\definecolor{longqual-3}{RGB}{51,160,44}
\definecolor{longqual-4}{RGB}{251,154,153}
\definecolor{longqual-5}{RGB}{227,26,28}
\definecolor{longqual-6}{RGB}{253,191,111}
\definecolor{longqual-7}{RGB}{255,127,0}
\definecolor{longqual-8}{RGB}{202,178,214}
\definecolor{longqual-9}{RGB}{106,61,154}
\definecolor{longqual-10}{RGB}{255,255,153}
\definecolor{longqual-11}{RGB}{177,89,40}

\title{Microarchitectural comparison and in-core modeling of state-of-the-art CPUs:\\ Grace, Sapphire Rapids, and Genoa
}

\author{\IEEEauthorblockN{Jan Laukemann}
\IEEEauthorblockA{\textit{Erlangen National High Performance Computing Center} \\
\textit{Friedrich-Alexander-Universität Erlangen-Nürnberg}\\
Erlangen, Germany \\
jan.laukemann@fau.de}
\and
\IEEEauthorblockN{Georg Hager}
\IEEEauthorblockA{\textit{Erlangen National High Performance Computing Center}\\
\textit{Friedrich-Alexander-Universität Erlangen-Nürnberg}\\
Erlangen, Germany \\
georg.hager@fau.de}
\and
\IEEEauthorblockN{Gerhard Wellein}
\IEEEauthorblockA{\textit{Erlangen National High Performance Computing Center}\\
\textit{Friedrich-Alexander-Universität Erlangen-Nürnberg}\\
Erlangen, Germany \\
gerhard.wellein@fau.de}
}

\maketitle

\begin{abstract}
With Nvidia's release of the Grace Superchip, all three big semiconductor companies in HPC (AMD, Intel, Nvidia) are currently competing in the race for the best CPU.
In this work we analyze the performance of these state-of-the-art CPUs and create an accurate in-core performance model for their microarchitectures Zen~4, Golden Cove, and Neoverse V2, extending the Open Source Architecture Code Analyzer (OSACA) tool and comparing it with LLVM-MCA.
Starting from the peculiarities and up- and downsides of a single core, we extend our comparison by a variety of microbenchmarks and the capabilities of a full node. The ``write-allocate (WA) evasion'' feature, which can automatically reduce the memory traffic caused by write misses, receives special attention; we show that the Grace Superchip has a next-to-optimal implementation of WA evasion, and that the only way to avoid write allocates on Zen~4 is the explicit use of non-temporal stores.
\end{abstract}

\begin{IEEEkeywords}
Intel Sapphire Rapids, NVIDIA Grace CPU Superchip, AMD Genoa, Golden Cove, Neoverse V2, Zen~4, in-core, performance analysis, performance modeling
\end{IEEEkeywords}


\section{Introduction}
\subsection{Motivation} 
The Grace Hopper Superchip as well as the Grace CPU Superchip~(GCS) mark the first HPC and data center systems by Nvidia with their own CPU, based on Arm's Neoverse V2 design.
One chip comprises 72 cores running at 3.4\,\GHZ\ all within one ccNUMA domain.
With this approach, Nvidia wants to catch up with the x86 competition and offer a full solution covering both accelerators (i.e., GPGPUs) and hosts (i.e., CPUs).
In this work, we analyze the Nvidia Grace CPU Superchip, compare its performance to the state-of-the-art competitor x86 CPUs Intel Sapphire Rapids~(SPR) and AMD Genoa, and provide an in-core performance model for all three microarchitectures for lower-bound runtime prediction that can be used as part of holistic performance models such as Roof{}line~\cite{Williams2009}, leading to valuable insights into performance bottlenecks of these new CPU architectures.

\subsection{Brief overview of the in-core port models}
When thinking about the performance of a single CPU core, we assume what is widely known as the \emph{port model}:
Each instruction, optionally split into one or more \emph{micro-ops}~(\muops), gets assigned to and executed by \emph{functional units}~(FUs) and may even require multiple FUs (e.g., to load data and do an arithmetic computation on the loaded value).
On the other hand, one FU might exist multiple times and can thus increase the instruction throughput via out-of-order~(OoO) execution; e.g., two FMA units that can be accessed in parallel double the throughput for this instruction type.
One or more FUs are grouped behind a \emph{port} as seen by the scheduler, i.e., for each port and each cycle, one \muop\ can be issued (with a global maximum number of \muops\ issued per cycle).
Figure~\ref{fig:v2-port-model} shows the port model of the Neoverse V2 microarchitecture, used in Nvidia's Grace CPU.
While the width of the SVE registers is relatively small (128\,bit), there is considerable instruction level parallelism~(ILP) in available ports with similar functional units.
As often seen in modern OoO-architectures, the execution of instructions including floating-point data is separated from the execution of integer data.
For more information about the idea of port models, see~\cite{Laukemann2018}.

\begin{figure}[tb]\centering
\includegraphics*[width=0.9\linewidth]{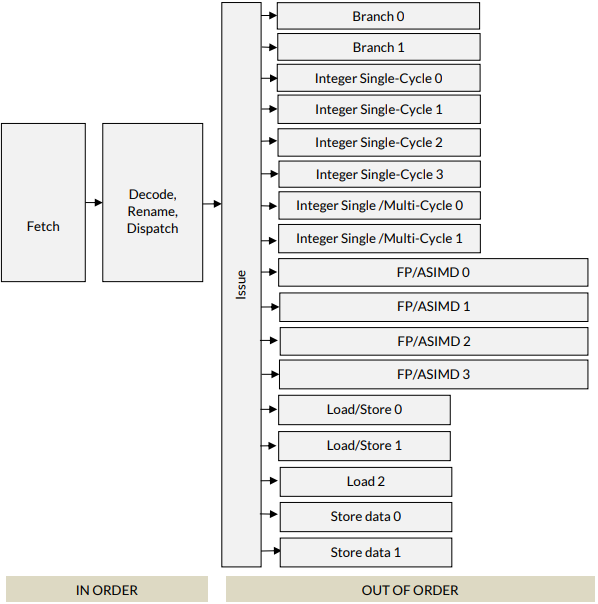}
\caption{Arm Neoverse V2 core block diagram and port model, compiled from Arm's Software Optimization Guide~\cite{NeoverseSOG}.
\label{fig:v2-port-model}}
\vspace{-0.5cm}
\end{figure}

\subsection{Testbed and experimental methodology} 
All experiments were carried out on dedicated servers in our test cluster: A two-socket Nvidia Grace CPU Superchip, a two-socket Intel Xeon Platinum 8470, and a two-socket AMD EPYC 9684X system.
The specific hardware features are listed in Table~\ref{tab:chipcompare}.
\begin{table*}[tb]
\centering\small
 \begin{tabular}{c | c c c} 
  & Nvidia Grace Superchip ``GCS'' & Intel Xeon Platinum 8470 ``SPR'' & AMD EPYC 9684X ``Genoa'' \\
 \hline\hline
 Cores                      & 72 & 52 & 96\\ 
 Frequency (max/base)     & 3.4\,\GHZ\ / 3.4\,\GHZ & 3.8\,\GHZ\ / 2.0\,\GHZ & 3.7\,\GHZ\ / 2.55\,\GHZ\\
 Theor. DP Peak  & 3.92\, Tflop/s & 6.32\,Tflop/s & 8.52\,TFlop/s \\
 Achiev. DP Peak  & 3.82\, Tflop/s & 3.49\,Tflop/s & 5.1\,TFlop/s \\
 TDP         & 250\,W & 350\,W & 400\,W\\
 Cache size (L1/L2/L3)      & 64\,KB / 1\,MB / 114\,MB & 48\,KB / 2\,MB / 105\,MB & 32\,KB / 1\,MB / 1152\,MB \\
 Main memory              & 240\,GB LPDDR5X & 512\,GB DDR5 & 384\,GB DDR5\\
 ccNUMA domains             & 1 & 4 (SNC-mode) & 1 (up to 4 configurable) \\
 Max. mem bandwidth    & 546\,GB/s / 467\,GB/s & 307\,GB/s / 273\,GB/s & 461\,GB/s / 360\,GB/s\\
 (theor. / measured) & & &\\
 \end{tabular}
 \caption{Comparison of the core features of the Grace CPU Superchip (GCS), the Intel Xeon Platinum 8470 (SPR), and the AMD EPYC 9684X (Genoa). For all servers, the L1 and L2 cache are exclusive caches per core, while the L3 is shared within one chip.}
 \label{tab:chipcompare}
 \vspace{-0.2cm} 
\end{table*}
For compilation we used GCC~12.1, the oneAPI~2023.2 compiler framework and LLVM Clang~17.0.6 for the x86 machines and the Arm C Compiler~23.10 (based on LLVM 17) and GCC~13.2 for the Grace server.
For cycle-accurate measurements we set the clock frequency to the 
corresponding base frequency using SLURM~\cite{Yoo2003} if possible, i.e., 2.0\,\GHZ\ for SPR and 2.55\,\GHZ\ for Genoa.
While Grace does not allow frequency fixing, we could not observe any frequency change running our benchmarks and validated the clock frequency of all runs with hardware performance counters using LIKWID~\cite{Likwid} 5.3.0~\cite{Likwid530}\footnote{As there is no official release of the LIKWID software with Nvidia Grace support and only limited support for Genoa, we used the development versions from \hyperlink{https://github.com/RRZE-HPC/likwid/pull/585}{PR585} and \hyperlink{https://github.com/RRZE-HPC/likwid/pull/618}{PR618}, respectively.}.
This tool was also used for all other hardware performance counter measurements.
To validate our in-core performance models, we used the OSACA\cite{Laukemann2018, Laukemann2019} tool in version 0.5.3 including our own extensions (which will be part of the next release) for supporting the microarchitectures in this paper.
Furthermore, the LLVM Machine Code Analyzer (LLVM-MCA)~\cite{llvm-mca} used for comparing the accuracy of our model.

\section{Architectural Analysis}
Since a port model visualization of all three microarchitectures as shown in Figure~\ref{fig:v2-port-model} would go beyond the scope of this work, we show the key aspects of the three cores in Table~\ref{tab:corecompare}.

\begin{table}[tb]
\centering\small
 \begin{tabular}{c | c c c} 
  & GCS &  SPR & Genoa \\
  & (Neoverse V2) & (Golden Cove) & (Zen~4) \\
 \hline\hline
 Number of ports    & 17    & 12    & 13 \\ 
 SIMD-width         & 16\,B & 64\,B & 32\,B \\
 Int units          & 6\tablefootnote{GCS's 6 Int ports comprise 2 multi-cycle + 4 single-cycle ports.} & 5 & 4\\
 FP vector units    & 4 & 3 & 4 \\
 Loads/cy           & $3 \times 128\,\textrm{B}$ & $2 \times 512\,\textrm{B}$ & $2 \times 256\,\textrm{B}$ \\
 Stores/cy          & $2 \times 128\,\textrm{B}$ & $2 \times 256\,\textrm{B}$ & $1 \times 256\,\textrm{B}$ \\
 \end{tabular}
 \caption{Comparison of the in-core features and port models for the GCS, SPR, and Genoa cores.}
 \label{tab:corecompare}
 \vspace{-0.5cm} 
\end{table}

While Golden Cove and Zen~4 have approximately the same number of ports (12 and 13, respectively), the Neoverse V2 stands out with its 17 ports, fully off{}loading any non-floating-point operations to other ports and providing a high ILP.
As a downside, even though the core supports the SVE vector extension for width-agnostic vector registers, the maximum register width is 128\,bit, which is only a fourth of Golden Cove's 512\,bit registers.
This leads to the expectation that the Golden Cove architecture can show its strength when executing highly vectorized code while the Neoverse V2 shines with code that is hard to vectorize and has many scalar instructions, as often seen in data center and AI workloads.
The Zen~4 meets the two extremes in the middle with 256-bit registers and slightly more ILP than Golden Cove.
Even though Zen~4 supports the AVX-512 extensions, their execution is split into $2 \times 256$\,bit packets.
While a comparison of the sustained peak memory bandwidths heavily depends on the built-in memory type and the number of DIMMs and would not represent a fair competition, we can compare them with the theoretical maximum and state that Genoa only achieves 78\% of its theoretical memory bandwidth peak, while GCS and SPR reach 87\% and 90\%, respectively.

While there is some documentation on the microarchitectures' backends~\cite{NeoverseSOG, intel_aorm, uopsinfo}, the information often is incomplete or insufficient to build a useful performance model.
Therefore, we write microbenchmarks with various benchmark tools~\cite{asmbench, ibench} for every interesting instruction to obtain its throughput, latency, and port occupation.
For the latter, it is often necessary to interleave the instruction with known instructions to infer the potential ports of execution.
While each model comprises hundreds of entries of individual combinations of assembly instructions and operands, we show the throughput and latency for some of the most important double-precision instructions in Table~\ref{tab:incore-tp-lt}.

\begin{table*}[tb]
\centering\small
 \begin{tabular}{c | c c c | c c c} 
  & GCS &  SPR & Genoa & GCS &  SPR & Genoa \\
  & (Neoverse V2) & (Golden Cove) & (Zen~4) & (Neoverse V2) & (Golden Cove) & (Zen~4) \\
  \hline
  Instruction & \multicolumn{3}{c|}{Throughput [DP elements / cy]} & \multicolumn{3}{c}{Latency [cy]} \\
 \hline\hline
 gather [CL/cy]     & \sfrac{1}{4}  & \sfrac{1}{3}  & \sfrac{1}{8}  & 9     & 20 & 13 \\
 VEC ADD            & 8             & 16            & 8             & 2     &  2 &  3 \\
 VEC MUL            & 8             & 16            & 8             & 3     &  4 &  3 \\
 VEC FMA            & 8             & 16            & 8             & 4     &  4 &  4 \\
 VEC FP Div         & 0.4           & 0.5           & 0.8           & 5     & 14 & 13 \\
 Scalar ADD         & 4             & 2             & 2             & 2     &  2 &  3 \\
 Scalar MUL         & 4             & 2             & 2             & 3     &  4 &  3 \\
 Scalar FMA         & 4             & 2             & 2             & 4     &  5 &  4 \\
 Scalar Div         & 0.4           & 0.25          & 0.2           & 12    & 14 & 13 \\
 \end{tabular}
 \caption{Throughput and latency for some double-precision instructions on GCS, SPR, and Genoa. If multiple values for one instruction were applicable, e.g., due to different performance for different vector widths, the best performance (i.e., highest throughput, lowest latency) was selected. Note that the throughput of the ``gather'' as a load instruction is given in ``cache lines per cycle'' while the rest is evaluated in double precision elements per cycle.}
 \label{tab:incore-tp-lt}
 \vspace{-0.2cm} 
\end{table*}


The Golden Cove architecture shows the highest throughput for all shown vector instructions due to its large register width. The Neoverse V2 can demonstrate its strength for scalar instructions due to the large ILP.
A single Zen~4 core is slower or breaks even in terms of throughput for all instructions shown; however, a full chip comprises 96 cores while an SPR chip comes only with 52 cores.
Therefore, if an application allows a high parallelism on the node level, e.g., through OpenMP, the overall throughput of the Genoa system might come out first, as shown in the artificial peak FLOP benchmark used in Table~\ref{tab:chipcompare}.
When looking at the latencies of the investigated instructions, one can clearly observe the superiority of the Neoverse V2 which shows a lower or even latency for every single instruction in Table~\ref{tab:corecompare}.
Thus, arithmetic-heavy latency-bound codes such as iterative solvers using the Gauss-Seidel method~\cite{Hofmann_2020} could benefit from running on a Grace CPU Superchip compared to the two competitors.
Especially Intel seems to trade off their high throughput performance against a relatively high instruction latency, even though they managed to decrease the ADD latency by half compared to the predecessor Ice Lake microarchitecture.

Although wide registers can provide a good out-of-the-box speedup when using vectorization on a single core, it is a well-known problem for Intel to require the cores to throttle down for AVX-512-heavy code and when using multiple cores due to thermal constraints.
Therefore we analyzed the sustained clock frequency for arithmetic-heavy codes while scaling across a socket on all systems (see Fig.~\ref{fig:freqs}).
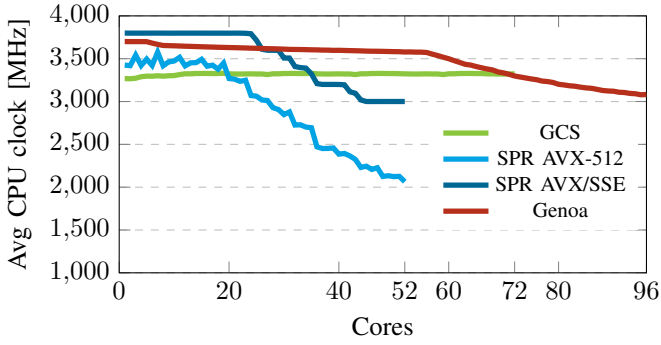
\begin{figure}[t]
	\centering
	\begin{tikzpicture}
    \begin{axis}[
        height=5cm,
        width=0.97\linewidth,
        xlabel={Cores},
        ylabel={Avg CPU clock [MHz]},
        xmin=0, xmax=96,
        ymin=1000, ymax=4000,
        xtick={0,20,40,52,60,72,80,96},
        ytick={1000, 1500, 2000, 2500, 3000, 3500, 4000},
        legend style={nodes={scale=0.8, transform shape}, cells={align=left}, draw=none, legend columns=1, at={(axis cs:57,1500)}, anchor=south west},
        ymajorgrids=true,
        grid style=dashed,
    ]
    \addplot[color=LimeGreen, line width=0.7mm]
        table[x expr=\thisrowno{0}, y expr=\thisrowno{1}, header=true, col sep=comma] {data/freq_GCS.csv};
        \addlegendentry{GCS}
    \addplot[color=Cerulean, line width=0.7mm]
        table[x expr=\thisrowno{0}, y expr=\thisrowno{1}, header=true, col sep=comma] {data/freq_spr.csv};
        \addlegendentry{SPR AVX-512}
    \addplot[color=MidnightBlue, line width=0.7mm]
        table[x expr=\thisrowno{0}, y expr=\thisrowno{2}, header=true, col sep=comma] {data/freq_spr.csv};
        \addlegendentry{SPR AVX/SSE}
    \addplot[color=BrickRed, line width=0.7mm]
        table[x expr=\thisrowno{0}, y expr=\thisrowno{1}, header=true, col sep=comma] {data/freq_genoa.csv};
        \addlegendentry{Genoa}
    \end{axis}
    \end{tikzpicture}
	\caption{Sustained CPU clock frequency for arithmetic-heavy code on GCS, SPR, and Genoa across one chip. If no ISA extension is specified, the architecture could sustain the same frequency for all supported ISA extensions.}
	\label{fig:freqs}
    \vspace{-0.5cm} 
\end{figure}
Each benchmark ran for several minutes and the clock frequency of all active cores was tracked using hardware performance counters.
While SPR shows a different behavior right from the start for AVX-512-heavy code, the sustained frequency for the GCS and Genoa did not change across ISA extensions.
Both SPR and Genoa eventually fall down to a frequency of 2.0\,\GHZ\ and 3.1\,\GHZ\ for AVX-512-heavy code, which results in 53\% and 84\% of their respective single-core turbo limit, even though SPR manages to sustain a frequency of 3.0\,\GHZ\ for the case of AVX- or SSE-heavy code (78\% of Turbo).
The Nvidia GCS exhibits a constant frequency of 3.4\,\GHZ\ (the base frequency) throughout the whole socket.
Therefore, for highly parallel arithmetic-heavy code, one might see better performance on GCS compared to SPR despite even for throughput- or latency-bound code due to a 1.7$\times$ higher sustained clock frequency. 

The individual measurements can be incorporated with the specific port occupations into an in-core performance model that can be used for optimistic runtime prediction or as a building block for node-wide performance models (e.g., a more realistic horizontal ceiling in the Roof{}line Model~\cite{Williams2009} or the in-core component of the Execution-Cache-Memory~(ECM) Model~\cite{Stengel2015}).
While there exists a wide range of tools capable of applying such a model automatically to a given code without compiling or running it~\cite{uica, llvm-mca, cqa, facile, granite} via \emph{static code analysis}, we choose to use the Open Source Architecture Code Analyzer (OSACA)~\cite{Laukemann2018, Laukemann2019} as it provides the user with the possibility of adding new microarchitectures into the existing framework relatively easily.
For validation of our models we used 13 streaming microbenchmarks (Jacobi [2D 5-point$|$3D 27-point$|$3D 7-point$|$3D 11-point] stencil, ADD, COPY, Gauss-Seidel 2D 5-point stencil, $\pi$-computation by integration, INIT, Schönauer Triad, Sum reduction, STREAM Triad~\cite{McCalpin1995}, UPDATE), compiled with different compilers (Armclang, GCC, oneAPI, and Clang) and different optimization flags (\texttt{-O1}, \texttt{-O2}, \texttt{-O3}, and \texttt{-Ofast}), resulting in 416 tests and 290 unique assembly representations.

\begin{figure}[tb]
	\centering
    \includegraphics[width=\linewidth]{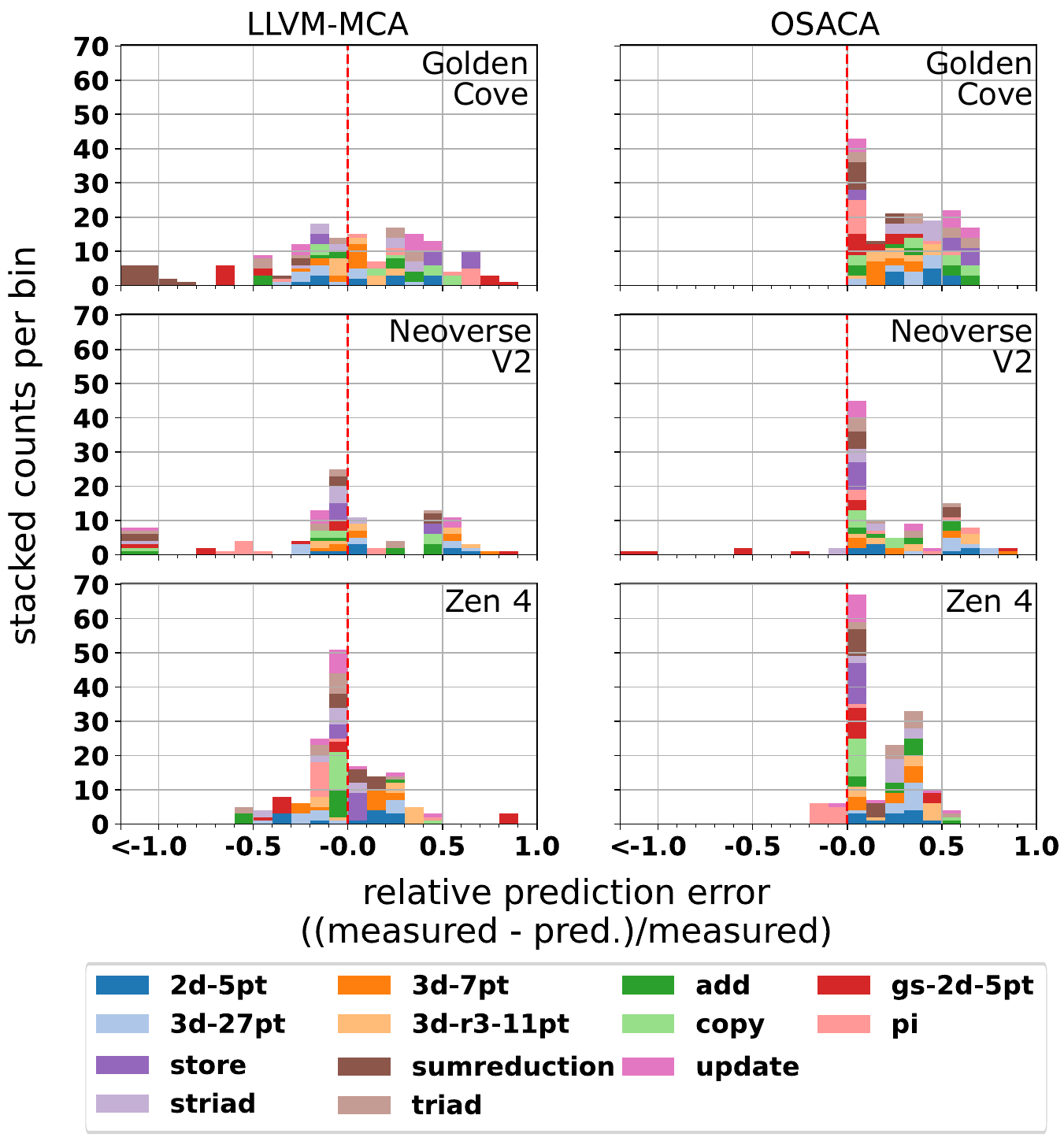}
	\caption{Relative prediction error of 416 test blocks for LLVM-MCA and OSACA. Bars right of the red dotted line indicate a prediction faster than the actual measurement while bars left of the line indicate a slower prediction. 
    }
	\label{fig:error-archs}
    \vspace{-0.5cm}
\end{figure} 

Figure~\ref{fig:error-archs} (based on graphs in~\cite{Hammer2023}) shows histograms of the relative prediction error~(RPE) for the kernels with our models of the investigated microarchitectures incorporated into OSACA versus the LLVM performance models in LLVM-MCA.
Each bucket marks a range of 10\% relative error; bars right of the red dotted zero line indicate a prediction faster than the actual measurement while bars left of the line indicate a slower prediction.
The bucket in the very left collects all predictions larger than -1.0 (i.e., off
by more than a factor of 2).
As we aim to provide a lower-bound estimate, we prefer to see all errors on the right of the zero line.
Except for a few versions of the Gauss-Seidel kernel for the Neoverse V2, where OSACA (correctly) predicts a register dependency that the CPU can overcome by register renaming, and the $\pi$ kernel for Zen~4, where our model assumes a lower throughput for the scalar divide than we measure, this is the case for all other tests (96\%) with our performance model.
There is one kernel predicted incorrectly by more than a factor of 2, and 37\% (44\%) are predicted accurately with a positive RPE of less than 10\% (20\%).
The LLVM-MCA model, however, predicts 75\% of the test kernels slower than the actual measurements, with 14 measurements being off by more than a factor of 2.
Only 10\% (16\%) are predicted correctly with positive a RPE of less than 10\% (20\%), although this value increases to 32\% (48\%) when considering the 10\% (20\%) bucket on the negative side of the zero line.
The average RPE of only the under-predictions (i.e., right-hand-side errors) of our model in OSACA shows a smaller error for Golden Cove, V2, and Zen 4 with 24\%, 30\%, and 18\% versus the LLVM model showing 38\%, 34\%, and 20\%.
When looking at the global (i.e., absolute) RPE, our model still performs better for Golden Cove~(30\% vs 35\%) and V2~(26\% vs 52\%), and is slightly worse for ZEN4 than the LLVM-based model~(18\% vs 16\%).


\section{Case Study: Write-Allocate Evasion}

One interesting feature that has entered x86 processors with the Intel Ice Lake generation is the automatic evasion of write-allocate (WA) transfers from memory. Write-allocate usually occurs in cache-based architectures when a standard store operation from a register to memory causes a write miss: Since the core can only communicate with its L1 cache, the cache line must be read from memory before it can be modified and then (later) written back. This extra data traffic can impact the performance and clutter the cache with data that may not be needed soon. \emph{Cache line claim} and \emph{non-temporal stores} are two ways to avoid write-allocates. Both can be supported by special instructions that claim a cache line in the cache without reading it first (available on some Arm CPUs) or write data to memory through a special write-combine buffer outside the normal cache hierarchy (available on Arm and x86 CPUs). Cache line claim can also be automatic if a core is able to detect that a cache line will be overwritten entirely. This feature has been supported for a long time by many Arm CPUs (including, e.g., the Marvell ThunderX2 and GCS) and by Intel server chips starting with the Ice Lake family, where Intel termed it \emph{SpecI2M}~\cite{papazian2020new}.

In order to fathom the ability of the CPUs under investigation to employ automatic and explicit WA evasion, we run a simple store-only (array initialization) benchmark, measure the actual memory data traffic (which includes write-allocates), and divide it by the amount of stored data. With perfect WA evasion in place, this ratio should be equal to one. It should be equal to two if the full WA transfers apply. On Intel CPUs it was shown previously~\cite{Laukemann2024,HagerSpecI2M} that the efficiency of SpecI2M depends crucially on the saturation of the memory interface: Only if a significant fraction of the maximum memory bandwidth is utilized will the WA evasion mechanism kick in. Figure~\ref{fig:stores} shows the results of the store benchmark with respect to the number of cores utilized for all three CPUs. In case of SPR and Genoa, we have added a variant with non-temporal (NT) stores for reference; ideally, NT stores should eliminate the WA transfers entirely. 

The results show that only GCS is able to completely avoid WA transfers automatically in this simple scenario (solid green line). The SpecI2M mechanism in SPR can only reduce write-allocates by up to 25\% and only kicks in when a large part of the 13 cores on a ccNUMA domain is utilized (solid blue line). The only way to WA evasion on Genoa is via non-temporal stores, which works perfectly, however (dotted dark red line vs.\ solid red line). Finally, on SPR even the non-temporal stores are not 100\% effective, and there is a  residual 10\%  of WA traffic except for very small core counts (dotted blue line).

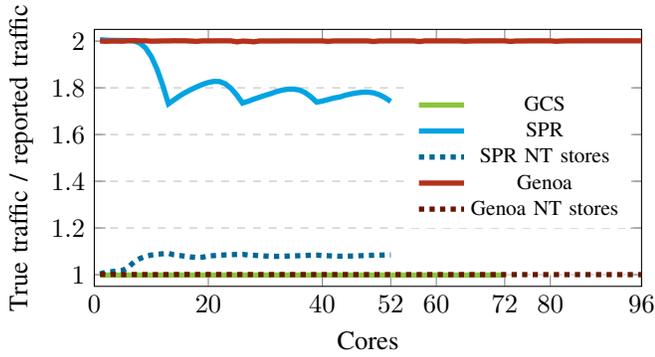
\begin{figure}[t]
	\centering
	\begin{tikzpicture}
    \begin{axis}[
        height=5cm,
        width=1.0\linewidth,
        xlabel={Cores},
        ylabel={True traffic / reported traffic},
        xmin=0, xmax=96,
        ymin=0.95, ymax=2.05,
        xtick={0,20,40,52,60,72,80,96},
        ytick={1.0, 1.2, 1.4, 1.6, 1.8, 2.0},
        legend style={nodes={scale=0.8, transform shape}, cells={align=left}, draw=none, legend columns=1, at={(axis cs:55,1.2)}, anchor=south west},
        ymajorgrids=true,
        grid style=dashed,
    ]
        \addplot[color=LimeGreen, line width=0.7mm]
        table[x expr=\thisrowno{0}, y expr=\thisrowno{1}, header=true] {data/wa-evasion.tsv};
        \addlegendentry{GCS}
    \addplot[color=Cerulean, line width=0.7mm]
        table[x expr=\thisrowno{0}, y expr=\thisrowno{2}, header=true] {data/wa-evasion.tsv};
        \addlegendentry{SPR}
    \addplot[color=MidnightBlue, line width=0.7mm, dotted]
        table[x expr=\thisrowno{0}, y expr=\thisrowno{3}, header=true] {data/wa-evasion.tsv};
        \addlegendentry{SPR NT stores}
    \addplot[color=BrickRed, line width=0.7mm]
        table[x expr=\thisrowno{0}, y expr=\thisrowno{1}, header=true] {data/wa-evasion-genoa.tsv};
        \addlegendentry{Genoa}
    \addplot[color=Sepia, line width=0.7mm, dotted]
        table[x expr=\thisrowno{0}, y expr=\thisrowno{2}, header=true] {data/wa-evasion-genoa.tsv};
        \addlegendentry{Genoa NT stores}
    \end{axis}
    \end{tikzpicture}
	\caption{Ratio of actual memory traffic to stored data volume vs.\ number of cores for a store-only benchmark loop (working set 40\,\GB). A value of $1.0$ indicates perfect WA evasion, while a value uf $2.0$ indicates full WA traffic. The variants labeled ``NT stores'' use non-temporal store instructions, while the others use standard stores. 
    }
	\label{fig:stores}
    \vspace{-0.5cm}
\end{figure}

\section{Conclusion}
Via a thorough in-core analysis of the Nvidia Grace CPU Superchip, the Intel Sapphire Rapids, and the AMD Genoa CPU, we showed peculiarities of their microarchitectures Neoverse V2, Golden Cove, and Zen~4, respectively, established an in-core performance model for each of them, and applied it to simple streaming kernels.
We showed that the models, incorporated in the Open Source Architecture Code Analyzer~(OSACA), yield  more accurate lower bounds for in-core runtime than the existing LLVM-MCA model for a comprehensive  set of microbenchmarks.
Furthermore, we investigated the node-level capabilities of the HPC servers such as the sustained CPU clock frequencies and the memory bandwidth and focused on implicit and explicit Write-Allocate evasion techniques.
In future work, we plan to continue these investigations by applying our in-core model to a node-wide performance model such as the Execution-Cache-Memory (ECM) model and study real-life applications on a larger scale.


\bibliographystyle{IEEEtran}
\bibliography{IEEEabrv,pmbs}

\end{document}